# Multipolar Reactive DPD: A Novel Tool for Spatially Resolved Systems Biology

Rudolf M. Füchslin*, Thomas Maeke, and John S. McCaskill[♦]

Ruhr-Universität Bochum, Biomolecular Information Processing (BioMIP),

c/o BMZ, Otto-Hahn-Str. 15, D-44227 Dortmund, Germany

*Corresponding author: rudolf.fuechslin@biomip.ruhr-uni-bochum.de

[♦]Currently on leave from Friedrich Schiller University, Jena, Germany

**Summary Sentence**: Chemical kinetics is augmented to include collective dynamics of extended structures (e.g. membranes) relevant to mesoscale cell simulations by means of a novel multipole reactive extension to dissipative particle dynamics.

**Abstract**

**Background:** The present advances in systems biology require a simulation platform that enables the study of the collective dynamics of complex chemical and structural systems in a spatially resolved manner with a combinatorially complex variety of different system constituents. In order to allow a direct link-up with experimental data (e.g. high throughput fluorescence images) the simulation platform must be constructed locally, which means that mesoscale phenomena have to emerge from local composition and interactions (chemical and physical) that can be extracted from experimental data, e.g. fluorescence images. Under suitable conditions, the simulation of such local interactions must lead to processes such as vesicle budding, transport of membrane bounded compartments and protein sorting, all of which result from a sophisticated interplay between chemical and mechanical processes and require the link-up of different length scales. This article reports about a novel extension of a well-established method that allows this goal to be achieved. **Methodology:** Dissipative particle dynamics (DPD) is a momentum conserving, coarse-grained particle based simulation method, which has been applied for the study of various soft-matter systems. We show that introducing multipolar interactions between particles leads to extended membrane structures emerging in a self-organized manner and exhibiting both the necessary mechanical stability for transport and fluidity so as to provide a two-dimensional self-organizing dynamic reaction environment for kinetic studies in the context of cell biology. We further show that the emergent dynamics of extended membrane bound objects is in accordance with scaling laws imposed by physics. . **Significance:** Employing the presented extension of DPD, processes connecting different length scales, ranging from that of chemical kinetics to the mesoscopic scale of cellular compartments are simulated in a way that allows a link-up with experimental high-throughput-imaging data and standard protocols of systems biology.

## Introduction

The recent advances in several fields, from theoretical aspects of self-assembly to the possibilities of live cell-imaging, are now calling for a new integration methodology beyond chemical kinetics, efficiently linking up theoretical models with experimental results by means of a versatile simulation platform. Two central aspects of chemical processing in living systems can be summarized by the key terms "combinatorial variability of the molecules involved" and "spatial organization and compartmentalization via self-organized, dynamic membrane structures". Though applicable to all organisms, the latter property has reached a much higher level of sophistication in eukaryotes than in prokaryotes: While prokaryotes have a highly dynamic plasma membrane that separates their interior from the environment, eukaryotes exhibit a complex system of internal membranes that form separated compartments communicating via vesicular traffic in a precisely orchestrated way. Additionally, the membranes themselves provide a two-dimensional reaction environment in which collective self-organization of embedded structures is utilized by the cell. A simulation method must reflect both aspects, topological and chemical membrane dynamics, whereby their combination poses a considerable challenge: On one hand, in order to serve as a chemical reaction environment, the membranes must be sufficiently fluid in order to allow diffusive transport, on the other hand, the membranes as a whole must remain structurally stable even under perturbations far beyond the thermodynamical level.

The interplay of local chemistry and mesoscale dynamics also raises issues of control. In man-made systems, the regulation of the topmost hierarchy level is usually done either according to a fixed protocol or by an external entity. This is not the case in biological cells, where mesoscale changes in the morphology and also topology of membrane-bound compartments are regulated on the molecular level in a self-organized manner that does not require a central control instance. To understand and quantify this type of evolved decentralized control and the interplay between local interactions and global structure and dynamics are core objectives of systems biology. This endeavor requires a simulation platform that is capable of representing chemical and morphological dynamics of a membrane system and is equipped with an interface to the standardized and accepted protocols of systems biology. This particularly entails compatibility with high-throughput data, which means that (ideally) all parameters of the platform should be automatically derived from experimental data such as fluorescence images and reaction kinetics databases. In practice, this excludes using non-local phenomenological rates of morphological changes as input parameters, if not achievable efficiently in an algorithmic manner. Note that this is not only a problem of image analysis but also of experimental capabilities, which often allow reasonable access to local data but may not be able to deliver genuine non-local quantities, especially not dynamical ones.

We illustrate this with an example: the budding (see Fig. 1, parameters and length scales are discsussed in the Methods and Results sections respectively) of transport vesicles from larger membrane compartments, followed by fusion with another compartment, is a basic mechanism for a variety of cellular processes including endo- and exocytosis. Protein sorting is intimately connected with these two cellular pathways [1,2,3]. One viable modeling strategy for such complex processes would seem to be to implement compartments as extended entities and to define fission/fusion processes with rate parameters. An alternative strategy, at a first glance much more cumbersome, implements compartments as composed of locally interacting entities which exhibit processes such as budding as a result of changes in local interaction parameters, e.g. molecular associations or reactions leading to a different local curvature of the membrane. Though being less direct from the point of view of the phenomena one wants to reproduce, and more demanding in terms of computational effort (the budding process itself is the result to be calculated from lower-level processes), this latter approach does allow a direct link-up of local data (e.g. from fluorescence imaging or known enzymatic reaction rates) with the mechanisms underlying the model. Note, however, that such a local approach can be (and in the context of this work is) still phenomenological and may have an intrinsic length scale significantly above the molecular level (while still being able to account for the diffusion of single molecular catalyst, for example). A simulation platform based on local interactions therefore should be capable of reproducing emergent properties on the mesoscale properly, but may not necessarily extend right down to molecular details of e.g. membrane structure.

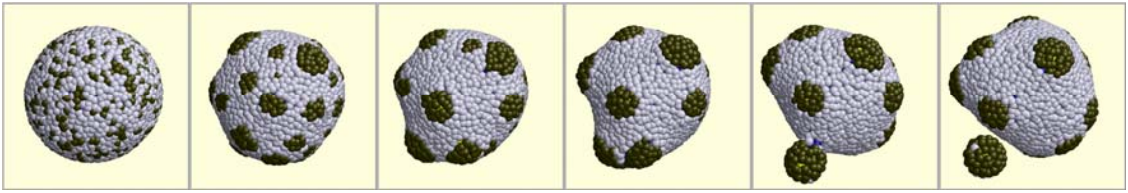

**Fig. 1: Budding of a small vesicle from a large compartment.** The figure illustrates the capability of the mprDPD-approach to model mesoscopic topological changes from local interactions. Starting from a spherical vesicle composed of two types of membrane particles, one observes the segregation with succeeding curvature induced budding. The initial vesicle contained 75% of type A and 25% of type B particles in random distribution. The interaction parameter where (only given if different from those in Table 1): $\kappa_{AA} = 0.3, \kappa_{BB} = 1.5, \alpha_{BB} = 110$ . For interactions between particles of different type, we set $X_{AB} = X_{BA} = (X_{AA} + X_{BB})/2$ for all parameters $X$ .

The peculiar features of living systems result in serious obstacles for implementing such a simulation platform. Several alternative approaches will be presented in the Discussion section 0 but they, though being very valuable for specific tasks, are hardly suitable for a generic platform that should allow the integrated simulation of the cellular machinery mentioned earlier.

In this article, we describe a simulation platform for self-organized, membrane based chemical processing of families of combinatorially diverse entities. As the underlying simulation method, we employ dissipative particle dynamics[4,5,6,7] (DPD), a coarse-grained particle based method that, though including stochastic interactions endowing the system with a temperature, conserves momentum and is therefore especially suited for investigations of the dynamics of extended objects composed of large number of individual entities. We extended classical DPD in a twofold manner: firstly by including chemical reactions and secondly by enabling an efficient self-organized treatment of supramolecular structures such as membrane bound compartments; the scope of this work is the latter, whereas the former will be presented in a subsequent article. Classical DPD studies the interaction of point-particles interacting by central forces. We extended this framework by equipping the DPD-particles with dipole moments and including interactions determined by a corresponding Lagrangian. These dipole moments have to be understood as abstract quantities expressing the first term in a multipole expansion and shall not be confused with the usual electrostatic dipole moments. The motivation behind this extension is the idea that a DPD-particle's dipole moment defines a local direction and can be understood as a surface element that can be used to build up extended curved two-dimensional objects embedded in space. Whereas this article is focused on the description of these extensions, demonstrating that the proper mesoscale dynamics emerges, a related article by the same authors presents applications of this work to problems in cell biology.

In recent years, various methods based on dissipative particle dynamics have been applied to the study of membranes and vesicles. Most of these implement membranes as being composed of polymer chains constructed from conventional DPD-particles. In order to get chains that exhibit some stiffness, Shillcock and Lipowski[8] introduced a bond-angle potential in their chains. Recently, polymersomes[9] or dynamic simulations of fusion events were performed using this method[10]. Jakobsen et. al. in a study of membrane fluidity[11] also used branched chains. Chain based methods proved to be very successful for studies close to the molecular lengths scale; however, in the context of endocytosis for example, for which the authors are endeavoring to present a physically grounded systems biology, the dynamics of whole membrane bound compartments (vesicles, endosomes at different stages of their maturation, etc.) needs to be simulated. We will show that the extended DPD-method we present enables the investigation of mesoscale phenomena in a stationary state as well as dynamical phenomena such as the correct hydrodynamic response of a vesicle to an external force. The latter is of relevance, because the response of membranes to localized external forces is a problem in the focus of present experimental investigations[12] that is direct relevance for endocytosis[13].

## Methods

### Classical Dissipative Particle Dynamics (DPD)

Dissipative particle dynamics is a now well-established method for the study of complex interacting systems, for a review, including a discussion of algorithmic efficiency, see[14]. A key motivation for its development was the need for a tool that enables studying soft matter systems at length scales above the computational limits of molecular dynamics, but retaining scaling properties lost by Brownian dynamics[15]. DPD combines three types of forces: conservative interactions, determining the macroscopic dynamics of extend objects, and dissipative and random forces that integrate the effects of molecular motion on faster timescales in a thermodynamically consistent way. There are several sophisticated integration schemes; the one employed in this work (the DPD-Velocity Verlet algorithm) is presented in Box 1, together with the definitions of all parameters and expressions employed. We will not discuss the algorithm in detail (for a thorough treatment, see[16]), but give a brief account of DPD and our rational for using it.

---

Box 1 The velocity-Verlet DPD Algorithm

The DPD-VV algorithm models the pairwise interaction of $N$ particles using three different types of central forces: $\vec{F}_C$ describes conservative (C) forces, $\vec{F}_R$ represents a random force (R) (and thereby establishes a temperature) and the dissipative (D) force $\vec{F}_D$ the corresponding (fluctuation-dissipation theorem) velocity attenuation. Because all these forces are central two-particles forces, momentum is strictly conserved. Furthermore, the interactions are of finite range, meaning that all forces vanish outside a cut-off radius $r_c$ . Some notational conventions are defined in the following. $\vec{F}_{X,ij}$ represents the force of type $X$ exerted on particle $j$ by particle $i$ , $\vec{F}_{X,j} = \sum_{i=1}^{N} \vec{F}_{X,ij}$ gives the total force of type $X$ on particle $j$ ( $X = C, R, D$ ). Accordingly, $\vec{r}_{ij} = \vec{r}_i - \vec{r}_j$ and the omission of the vector sign means automatically reference to the absolute value, e.g. $r_{ij} = \left| \vec{r}_{ij} \right|$ . The unit vector connecting particle $i$ and $j$ is defined as $\vec{e}_{ij} = \vec{r}_{ij} / r_{ij}$ .

The forces are given by the following expressions:

$$\vec{F}_{C,ij} = -\frac{\partial}{\partial r_{ij}} U_M(r_{ij}) \vec{e}_{ij}, \quad U_M(r) \equiv 0 \text{ for } r > r_C \tag{1}$$

$U_M$ stands here for the monopole potential. There is considerable freedom in choosing its precise form. Our choice is described and motivated in the text. For the random force, one employs

$$\vec{F}_{R,ij} = \sigma \omega^R(r_{ij}) \xi_{ij} \vec{e}_{ij} \tag{2}$$

with $\sigma$ a constant, $\omega^R(r_{ij})$ a function representing shielding and $\xi_{ij}$ a random variable (Gaussian white noise) with $\left\langle \xi_{ij} \right\rangle = 0$ and $\left\langle \xi_{ij}^2 \right\rangle = 1$ . Finally, the dissipation is given by

$$\vec{F}_{D,ij} = -\gamma \omega^D(r_{ij})(\vec{v}_{ij} \cdot \vec{e}_{ij}) \vec{e}_{ij} . \tag{3}$$

Again, $\gamma$ is a constant, $\vec{v}_{ij}$ the relative center-of-mass velocity of the particle and the dot denotes a scalar product.

In order to satisfy the fluctuation-dissipation theorem $\omega^D$ there $\omega^R$ must be related in the form:

$$\omega^D = (\omega^R)^2 . \tag{4}$$

In order to ensure a smooth decline of random forces as the cutoff radius is approached, one usually sets

$$\omega^R(r_{ij}) = \begin{cases} (1 - \dfrac{r_{ij}}{r_C}), & r_{ij} < r_C \\ 0, & r_{ij} \geq r_C \end{cases} . \tag{5}$$

Finally, fluctuation and dissipation are related via the temperature by

$$\sigma^2 = 2\gamma k_B T . \tag{6}$$

The integration algorithm (DPD-VV) is given by:

1.  $\vec{v}_i \leftarrow \vec{v}_i + \dfrac{1}{2m_i}(\vec{F}_{C,i}dt + \vec{F}_{D,i}dt + \vec{F}_{R,i}\sqrt{dt})$.

2.  $\vec{r}_i \leftarrow \vec{r}_i + \vec{v}_i dt$.

3.  Update $\vec{F}_{C,i}, \vec{F}_{D,i}, \vec{F}_{R,i}$.

4.  $\vec{v}_i \leftarrow \vec{v}_i + \dfrac{1}{2m_i}(\vec{F}_{C,i}dt + \vec{F}_{D,i}dt + \vec{F}_{R,i}\sqrt{dt})$.

5.  Update $\vec{F}_{D,i}$, go to 1.

(7)

The square root of the time increment for the integration of the random force results from the required stochastic integration of such forces[17].

The DPD approach can be summarized by noting that (i) one applies a coarse graining, which implies that an individual particle in a DPD simulation represents a large bunch of physical molecules, (ii) degrees of freedom lost by coarse graining are replaced by random noise, (iii) dissipation is added to compensate the resulting energy increases, by a force that is proportional to the center-of-mass velocity of the particles involved, (iv) all interactions, including the random and the dissipative one, are modeled by momentum conserving, central two-particle forces, and (v) the corresponding interaction potentials are set to zero outside a given cut-off radius $r_C$ and are soft in the sense that they exhibit no singularity for vanishing particle distance.

Momentum conserving, non-singular forces are peculiar to DPD and deserve some discussion. The strict momentum conservation specifically supports the simulation of composed, extended objects. This is because the conservation of linear momentum yields proper transfer of impulses across collective structures and hence allows directed motion of non-rigid, larger objects as in the motion of vesicles dragged along the cytoskeleton by motor proteins. Conservation of angular momentum is trivially given in classical DPD (two particle interactions mediated by central forces yield zero torque) but, in the extension of DPD we propose, the proper handling of torque turned out to be crucial. The fact that no real hard-core repulsive forces are employed at short distance between particles may on a first glance look strange; one is accustomed to forces which attain infinite values for particles with zero distance, as for example the forces resulting from Lennard-Jones potentials. However, this absence of singularities is appropriate since particles represent collections of physical molecules[6]. Limiting the magnitude and range of forces yield much more efficient simulations. Potential singularities demand a careful handling of the size of the time step for convergence, usually resulting in rather small time increments (on the femtosecond scale for MD), whereas limited forces allow the individual integration steps to be carried out over much larger time intervals.

**Dissipative Particle Dynamics with Multipole Interactions**

Conventional DPD is based on structureless point particles, *i.e.* particles that are completely determined by their position, apart from scalar properties such as mass. Such particles interact with their environment (i.e. with other particles) only through radially symmetric interactions, which means that the forces (whether conservative, dissipative or random) depend only on their mutual distance, $r_{ij}$, and show no further dependencies on position coordinates $\vec{F}_{X,ij}(\vec{r}_{ij}) = \vec{F}_{X,ij}(r_{ij})$. The problem is that, by tailoring such forces, it is not directly possible to self-organize extended flat structures, as would be necessary for the implementation of membranes for example. The reason is that the elements forming a surface have to be capable of relatively free movements within the surface but must resist movements that would lead to a strong bending of the surface. There are several ways to circumvent this problem: firstly, one could introduce higher-order multi-particle interactions that are equivalent to the evaluation of local curvature (net-models of membranes[18]). A second possibility involves the introduction of "chain-particles", *i.e.* DPD-particles that are connected with stiff springs and which exhibit a three-body angular potential between three successive particles in the chain, giving it sufficient stiffness to remain extended. This approach[8] yielded excellent results for small patches of membrane surfaces (it was even possible to study the internal stress distribution inside the membranes). However, the method is not easy to extend to larger length scales. This is not only because calculating the interaction between chains of particles is computationally costly, owing to the larger number of particles, but also because the introduction of stiff springs is numerically unstable at large integration steps. A third possibility, and the one we shall pursue here, is to replace the structureless point particles of conventional DPD by particles carrying additional multipole moments.

One could consider the chain-particles of Shillcock and Lipowski[8] above either as a collection of structureless DPD particles interacting in a specific manner (stiff springs, angular potential) or as one single super-particle composed of sub-units. These super-particles are no longer spherical and consequently their mutual interaction is not isotropic but depends on their relative orientation. This for example will allow some local control of surface bending when the surfaces are formed by these entities. A general asymmetric potential can be described by a multipole expansion in spherical harmonics. For the results presented in this work, it turned out that already a dipole moment is sufficient, but for other applications, also low order, higher multipole moments may be necessary. The term "multipole" has here to be taken in the strict mathematical sense of a multipole expansion and is independent of electrodynamic effects such as distributions of charges or magnetic fields. A dipole moment in the sense we use it here can be understood just as a vector defining a direction with a certain magnitude. In this work, we will demonstrate the self-assembly of flat and intrinsically curved membranes from a single layer of membrane particles, whereby such a particle represents a whole patch of membrane defined by its surface normal. This may appear to be in contradiction with the fact that biological membranes consist of a bilayer, but note that this is a question of the length scales one wants to study. In this work we aim at a length scale of whole membrane bound compartments and their behavior and dynamics as whole objects. A simulation at a finer level of coarse-graining can always be performed, thereby also resolving structural details of the membranes, but of course this increases the computation costs. The determination of the length scale one aims to work at is a non-trivial task and one has to keep in mind that a simulation method that is based on coarse graining has a natural upper bound for resolution (which is basically given by the size of the particles resulting from the coarse graining process). However, as will be briefly discussed, the concept of oriented DPD-particles can by employing adequate potentials easily be applied to the study bilayered structures, thereby enabling the investigation of amphiphilic system at smaller length scales (including e.g. micellar systems). The interpretation of a single particle then changes from a whole patch of membrane towards (small groups of) individual amphiphiles with a distinguished direction (head and tail). This interpretation establishes a direct link to the chain particles of[8].

The use of particles carrying dipole moments poses a new problem, namely the calculation of the dynamics of objects with additional degrees of freedom. In the case of simple point particles the dynamics is determined by simple forces, which in the case of the conservative force between two point particles is just given by the negative gradient of a potential that depends on their distance. Equilibrium positions are then determined by the minima of the potential under discussion, apart from entropic effects. The directional degrees of freedom given by the dipole moments require classical mechanics with generalized coordinates, which can be achieved by working in the Lagrangian formalism. The basic equations are:

$$\frac{d}{dt}\frac{\partial}{\partial \dot{q}_i}L - \frac{\partial}{\partial q_i}L = 0,$$
$$L = T - U$$
(8)

where $T(q_1,...q_n)$ is the kinetic energy, $U(q_1,...q_n)$ the potential energy and the $q_i$ are generalized coordinates.

The use of dipole moments implies that there are two different types of generalized coordinates: the usual positions and additionally two degrees of freedom determining the direction of the dipole moment. It is assumed that the absolute value of the dipole moment itself is a constant; this means that the dipole moment can be envisaged as an orientation vector that rotates but does not change its length (which we then set to one). With this interpretation, it becomes clear that the dynamics resulting from the eqs. (8) can be interpreted in terms of conventional forces and torques, which in turn implies that each particle, besides its velocity $\vec{v}_i$ and linear momentum $\vec{p}_i$ carries an angular velocity $\vec{\omega}_i$ and an angular momentum vector $\vec{L}_i$. Angular velocities are retained as explicit variables to allow correct torque transfer (The simultaneous usage of $L$ and $\vec{L}$ for the Lagrangian function and the angular momentum respectively, as well as $T$ and $\vec{T}$ for kinetic energy and torque may appear confusing but follows standard practice.)

It is important to note that, as in the case of potentials depending only on particle positions, the equilibrium states are, again apart from entropic effects and fluctuations, given by the minima of the potentials. This in turn means that in order to match a given experimental situation, one has to identify those local configurations of particle position and orientation necessary for constructing the extended structures one wants to study (here two dimensional, closed surfaces) and then construct a potential that is minimal for these configurations. To summarize, the membrane forming particles are represented as multipole expansions truncated at the dipole term and their interactions are governed by a Lagrangian that is chosen in such a way that it exhibits the same symmetries as the membranes one wants to study. The configurations we look for are given by dipole moments aligned in parallel and perpendicular to the lines connecting two particles. These connecting lines form a surface, when understood as a virtual mesh. As long as one only wants to produce flat surfaces, the requirement of parallel dipole moments could be skipped, because of the collective alignment of three of more dipoles, but the directionality of the surface becomes important as soon as one wishes to implement curved structures.

A potential with a minimum for a flat surface is given by

$$U(\vec{r}_i, \vec{d}_i, \vec{r}_j, \vec{d}_j) = U_M(\vec{r}_{ij}) + U_\perp(\vec{d}_i, \vec{d}_j, \vec{r}_{ij}) + U_\parallel(\vec{d}_i, \vec{d}_j),$$

$$\vec{r}_{ij} = \vec{r}_i - \vec{r}_j; \qquad \vec{e}_{ij} = \frac{\vec{r}_{ij}}{r_{ij}}, \qquad r_{ij} = \left| \vec{r}_{ij} \right|,$$

$$\Omega(r_{ij}) = \begin{cases} \dfrac{1}{2}\left(1 - \dfrac{r_{ij}}{r_C}\right)^2, & r_{ij} < r_C \\[2mm] 0, & r_{ij} \ge r_C \end{cases} \tag{9}$$

$$U_\perp(\vec{d}_i, \vec{d}_j, \vec{r}_{ij}) = \alpha \Omega(r_{ij})\left(\left(\vec{d}_i \cdot \vec{e}_{ij}\right)^2 + \left(\vec{d}_j \cdot \vec{e}_{ij}\right)^2\right), \quad \alpha \ge 0, \text{ prefers } \vec{d} \text{ perpendicular } \vec{e},$$

$$U_\parallel(\vec{d}_i, \vec{d}_j) = -\beta \Omega(r_{ij})(\vec{d}_i \cdot \vec{d}_j), \quad \beta \ge 0, \text{ prefers } \vec{d}_i \text{ and } \vec{d}_j \text{ parallel.}$$

Here $\vec{d}_i$ determines the dipole moment and $U_M(\vec{r}_i, \vec{r}_j)$ represents the isotropic potential (*i.e.* the usual monopolar potential) between the particles $i$ and $j$. This type of potential has already been investigated by Dawson and Kurtovic in a study of self-assembly of amphiphiles on a lattice[19].

The derivatives of the monopole potential $U_M(\vec{r}_{ij})$ yield the forces necessary for stabilizing the inter-particle distances. These forces are smooth (without any singularity) and vanish outside a cut-off radius $r_C$. At a first glance, it is surprising that for example in the case of water they are given just as linear functions of the form:

$$\vec{F}_{ij} = f(r_{ij})\vec{e}_{ij}, \quad f(r) = \begin{cases} F_{\max}(1 - \dfrac{r_{ij}}{r_C}), & r_{ij} < r_C \\[2mm] 0, & r_{ij} \ge r_C \end{cases} \tag{10}$$

Note that for $F_{\max} > 0$, this force is repulsive. This formulation is not only chosen for computational simplicity, but yields the correct thermodynamic behavior of water. Using methods from statistical mechanics, Groot and Warren[6] justified this choice for the force function and calculated e.g. the resulting compressibility as a function of the parameter $F_{\max}$. Based on these considerations, we choose the same function for the water interaction and introduced for the forces between membrane particles or between membrane particles and solvent a slightly more general functional form of the force allowing small range repulsion and intermediate range attraction (see Fig 2A):

$$f(r) = \begin{cases} F_{max}(1 - \dfrac{r}{r_{cf}}), & r < r_{cf} \\[2mm] -\dfrac{2F_{min}(r - r_{cf})}{r_c - r_{cf}}, & r \geq r_{cf}, \qquad r < \dfrac{r_{cf} + r_C}{2} \\[2mm] -F_{min}(1 - \dfrac{2r - r_C - r_{cf}}{r_C + r_{cf}}), & r \geq \dfrac{r_{cf} + r_C}{2}, \qquad r < r_C \\[2mm] 0 & r \geq r_C \end{cases} \qquad (11)$$

This force function depends basically on three parameters, namely the maximal repulsion $F_{max}$, the maximal attraction $F_{min}$ and the equilibrium distance $r_{cf}$ (for the parameters used in the simulation presented in this work consult Table 1.).

The fact that the potentials we use are not purely repulsive makes it difficult to stabilize the system against collapse. One possibility to avoid this problem is the introduction of density dependent forces, as pioneered for DPD by Pagonabarraga and Frenkel[20]. We follow closely their presentation, but take into account the discussion by Warren[21] and Trofimov et al.[22]. They defined for each particle a density by

$$n_i = \frac{15}{2\pi} \sum_{j \neq i, r_{ij} < 1} \left(1 - r_{ij}\right)^2 \qquad (12)$$

and an (excess free) energy by

$$U_{DF,i} = \sum_{j \neq i} \psi(n_j) \qquad (13)$$

with $\psi(n)$ a function of the density. The resulting forces are then given by

$$\vec{F}_i = -\sum_{j \neq i} \nabla_{\vec{r}_i} \psi(n_j) = -\sum_{j \neq i} \frac{15}{\pi} \left( \frac{\partial \psi(n_i)}{\partial n_i} + \frac{\partial \psi(n_j)}{\partial n_j} \right)(1 - r_{ij}) \vec{e}_{ij} \qquad (14)$$

Note that, as it is the custom throughout the literature, we here set $r_C = 1$, in order to avoid unnecessary complicated notations. It remains to determine $\psi$. Setting $\psi \sim n$ yields a Groot-Warren fluid[6]. The next higher order expression is given by $\psi(n) = bn^2$, the form that used in this work. Note that we already accounted for the linear term in the monopole-potential.

In order to describe curved surfaces, the potentials in eqs. (9) must be slightly modified. Instead of being perpendicular to the connection $\vec{r}_{ij}$ and parallel to each other, the dipoles $\vec{d}_i, \vec{d}_j$ shall include default angles $\phi_i^{def}$ and $\phi_j^{def}$ with $\vec{r}_{ij}$, respectively (see Fig. 2B). The included default angle $\phi_{ij}^{def}$ between the two dipoles is then given by $\phi_{ij}^{def} = \phi_i^{def} - \phi_j^{def}$ (not explicitly shown in Fig. 2B). For a spherical surface, the angle between two normals depends on their distance. Consequently, we have

$$\phi_i^{def} = \frac{\pi}{2} - \left( \frac{\kappa_0 + \kappa r_{ij}}{2} \right), \phi_j^{def} = \frac{\pi}{2} + \left( \frac{\kappa_0 + \kappa r_{ij}}{2} \right),$$

$$\phi_{ij}^{def} = \phi_j^{def} - \phi_i^{def} = (\kappa_0 + \kappa r_{ij}) \tag{15}$$

with $\kappa, \kappa_0$ the parameter defining the curvature of the membrane. In this work, $\kappa_0$ is always set to zero. A non-zero value proved to be adequate in the study of micelles, where the interpretation of a DPD-particle is closer to a single molecule; whereas $\kappa$ is a pure geometrical parameter, $\kappa_0$ reflects the head-to-tail diameter ratio of a lipid.

Elementary geometrical considerations show that potentials derived from the relative orientation of two normalized vectors $\vec{a}$ and $\vec{b}$ and exhibiting extrema for a general default angle $\phi_0$ can be expressed as functions of scalar products $(\vec{a} \cdot \vec{c})$ with $\vec{c} = \cos(\phi_0)\vec{b} + \sin(\phi_0)\vec{\Psi}$, whereby $\vec{\Psi} = \left( (\vec{b} \times \vec{a}) / \left| \vec{b} \times \vec{a} \right| \right) \times \vec{b}$.

The potential then reads (to emphasize the relation to eqs. (9), we use again $U_\perp$ and $U_\parallel$ though these potentials will no longer lead to a strictly perpendicular or parallel orientation of the vectors involved):

$$\vec{c}_i = \cos(\phi_i^{def})\vec{e}_{ij} + \sin(\phi_i^{def}) \frac{(\vec{e}_{ij} \times \vec{d}_i)}{\left| \vec{e}_{ij} \times \vec{d}_i \right|} \times \vec{e}_{ij},$$

$$\vec{c}_j = \cos(\phi_j^{def})\vec{e}_{ij} + \sin(\phi_j^{def}) \frac{(\vec{e}_{ij} \times \vec{d}_j)}{\left| \vec{e}_{ij} \times \vec{d}_j \right|} \times \vec{e}_{ij},$$

$$\vec{c}_{ij} = \cos(\phi_{ij}^{def})\vec{d}_j + \sin(\phi_{ij}^{def}) \frac{(\vec{d}_j \times \vec{d}_i)}{\left| \vec{d}_j \times \vec{d}_i \right|} \times \vec{d}_j,$$

$$U(\vec{r}_i, \vec{d}_i, \vec{r}_j, \vec{d}_j) = U_M(\vec{r}_{ij}) + U_\perp(\vec{d}_i, \vec{d}_j, \vec{r}_{ij}) + U_\parallel(\vec{d}_i, \vec{d}_j),$$

$$U_\perp(\vec{d}_i, \vec{d}_j, \vec{r}_{ij}) = -\alpha\Omega(r_{ij}) \left( \left( \vec{d}_i \cdot \vec{c}_i \right)^2 + \left( \vec{d}_j \cdot \vec{c}_j \right)^2 \right), \quad \alpha \ge 0, \tag{16}$$

$$U_\parallel(\vec{d}_i, \vec{d}_j) = -\beta\Omega(r_{ij})(\vec{d}_i \cdot \vec{c}_{ij}), \quad \beta \ge 0.$$

At a first glance, these expressions may look slightly asymmetrical in the two indices, especially in the term $U_\parallel$. Note, however, that the geometrical idea of this potential is that the scalar product of a rotated $\vec{d}_j$ with $\vec{d}_i$ is calculated. One could as well apply the adjoint rotation on $\vec{d}_i$ and evaluate the scalar product with $\vec{d}_j$. Furthermore, the sign in front of $U_\perp$ is changed. Instead of requiring the dipoles being perpendicular to $\vec{r}_{ij}$, the minimum of the potential is now attained if the dipoles $\vec{d}_k, k = i, j$ are parallel to a rotated version of $\vec{r}_{ij}$, namely $\vec{c}_k$, whereby $\vec{c}_k$ and $\vec{r}_{ij}$ include the respective default angle.

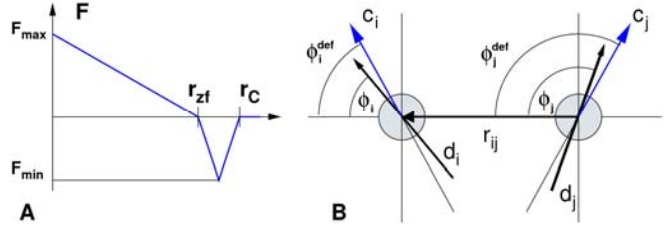

Fig. 2: A: **The monopole forces between two membrane particles.** Short range repulsion and intermediate range attraction lead, together with the dipole-interactions, to a stable membrane. B: Angular quantities used in the definition of the dipole-potentials for curved membranes.

Employing the Lagrangian formalism, one obtains for the dynamics (for the definition of the angles $\phi_i, \phi_j$ see Fig. 2B)

$$\frac{d\vec{p}_i}{dt} = \vec{F}_i^C = \sum_{j=1}^{N} \vec{F}_{ij}^C = -\sum_{j=1}^{N} \nabla_{\vec{r}_i} \left( U_M(\vec{r}_{ij}) + U_\perp(\vec{d}_i, \vec{d}_j, \vec{r}_{ij}) + U_\parallel(\vec{d}_i, \vec{d}_j) \right),$$

$$\frac{d\vec{L}_i}{dt} = \vec{T}_i^C = \sum_{j=1}^{N} (\vec{T}_{ij}^\perp + \vec{T}_{ij}^\parallel),$$

$$
\begin{aligned}
\nabla_{\vec{r}_i} U_\perp(\vec{d}_i, \vec{d}_j, \vec{r}_{ij}) = -\alpha \Big[ & \left( \cos(\phi_i - \phi_i^{def})^2 + \cos(\phi_j - \phi_j^{def})^2 \right) \nabla_{\vec{r}_i} \Omega(r_{ij}) \\
& + 2\sin(\phi_i - \phi_i^{def}) \cos(\phi_i - \phi_i^{def}) \Omega(r_{ij}) \nabla_{\vec{r}_i} \phi_i^{def} \\
& + 2\sin(\phi_j - \phi_j^{def}) \cos(\phi_j - \phi_j^{def}) \Omega(r_{ij}) \nabla_{\vec{r}_i} \phi_j^{def} \\
& - 2\sin(\phi_i - \phi_i^{def}) \cos(\phi_i - \phi_i^{def}) \frac{\Omega(r_{ij})}{r_{ij}} \frac{(\vec{e}_{ij} \times \vec{d}_i) \times \vec{e}_{ij}}{|\vec{e}_{ij} \times \vec{d}_i|} \\
& - 2\sin(\phi_j - \phi_j^{def}) \cos(\phi_j - \phi_j^{def}) \frac{\Omega(r_{ij})}{r_{ij}} \frac{(\vec{e}_{ij} \times \vec{d}_j) \times \vec{e}_{ij}}{|\vec{e}_{ij} \times \vec{d}_j|} \Big]
\end{aligned}
$$

$$\nabla_{\vec{r}_i} U_\parallel(\vec{d}_i, \vec{d}_j) = -\beta \Big[ \cos((\phi_j - \phi_i) - \phi_{ij}^{def}) \nabla_{\vec{r}_i} \Omega(r_{ij}) + \sin(\phi_i - \phi_i^{def}) \Omega(r_{ij}) \nabla_{\vec{r}_i} \phi_i^{def}$$

$$\vec{T}_{ij}^\perp = -2\alpha \left[ \cos(\phi_i - \phi_i^{def}) \sin(\phi_i - \phi_i^{def}) \frac{\vec{e}_{ij} \times \vec{d}_i}{|\vec{e}_{ij} \times \vec{d}_i|} + \cos(\phi_j - \phi_j^{def}) \sin(\phi_j - \phi_j^{def}) \frac{\vec{e}_{ij} \times \vec{d}_j}{|\vec{e}_{ij} \times \vec{d}_j|} \right], \tag{17}$$

$$\vec{T}_{ij}^\parallel = \beta \sin((\phi_j - \phi_i) - \phi_{ij}^{def}) \frac{\vec{d}_i \times \vec{d}_j}{|\vec{d}_i \times \vec{d}_j|}, \; .$$

with $\vec{T}_i$ representing torque. Of course, Newton's 3rd law is observed $\vec{F}_{ij} = -\vec{F}_{ji}$ and $\vec{T}_{ij} = -\vec{T}_{ji}$.

Up to now, we investigated only conservative forces: it remains to discuss the effect of fluctuation and dissipation on the dipoles. Based on Espanol and Warren[5], we used:

$$\vec{T}_{ij}^D = \gamma_D \omega^D(r_{ij})(\vec{e}_{ij}(\vec{e}_{ij} \cdot (\vec{\omega}_j - \vec{\omega}_i)))$$
$$\vec{T}_{ij}^R = \sigma \omega^R(r_{ij})\vec{n}_{ij}\xi_{ij}$$

(18)

with $\vec{\omega}_i$ the angular velocity of particle $i$. As will be shown in the result section, this dissipative and random torque in fact leads to the proper equipartition of energy.

Finally, we are investigating self-assembling systems with anisotropic interactions. This means that starting from a random initial configuration, the equilibration of the system may well lead to a more optimal packing, with other words a decrease of internal pressure. In order to compensate for this, we introduced a barostate. The pressure is calculated using the virial theorem; as shown in[6], it is sufficient to consider conservative forces means

$$p = \rho k_B T + \frac{1}{3V}\left\langle \sum_{j>i} \vec{r}_{ij}\vec{F}_{ij}^C \right\rangle$$

(19)

whereby the bracket denote a time average. The pressure $p$ was periodically measured and if a deviation of more than 2% from the expected pressure for water was detected, water particles were removed from or added to the system.

## Results

The concept of mprDPD as presented in the preceeding section is now shown to lead to membrane bound compartments exhibiting the correct emergent dynamics. Firstly, length and time scales and the self-assembly of different phases are discussed. Secondly, the statistical mechanics of an extended object is tested. A membrane bound compartment is placed in a harmonic spherical potential and its trajectory, resulting from random fluctuation is analyzed and compared with respect to expectations from equilibrium statistical mechanics. Thirdly, vesicles of variable diameter are pulled with a constant force and their stationary drift velocity is compared to those values predicted by Stokes' law. These tests show that, besides the ability to form membrane bound structures as such, mprDPD also reproduces correctly relevant emergent mesoscale dynamics and inhomogeneous kinetics and therefore opens up a wide range of applications of the method.

### Scales and Self-Assembly and Thermalization

DPD is a mesoscopic simulation method with inherently free time scale. This means that assuming a physical value for the cut-off radius $r_C$, the time scale has to be gauged by comparing with physical constants. One possibility is to evaluate the in-plane diffusion of DPD-particles forming flat membranes[11].

Assuming the physical size of $r_C = 100$ nm and a in-plane diffusion constant of $D \approx 4 \cdot 10^{12}$ m²/s, we get a unit of time of $\tau = 10^{-4}$ s. This may look surprisingly large: Though originally intended as a mesoscopic simulation method, DPD and consequently mprDPD, were assumed to behave badly under upscaling due to an unfavorable scaling behavior of the conservative interaction parameter (see e.g. [23]). We, however, claim DPD to be scalable assuming an appropriate renormalization procedure[24]. The simulation parameters we use are, if not stated differently, those given in Table 1; they are chosen to be roughly comparable to those used by Yamamoto and Hyodo[25].

In Fig. 3 we present self-assembled phases, showing the ability of mprDPD to reproduce at least important aspects of the phase diagram of amphiphilic systems. Also of relevance in our context is that for the simulation of system with about 21000 particles and a time step of $\Delta t = 3 \cdot 10^{-3}$, we get perfect equipartition between rotational and translational degrees of freedom with energy fluctuations lower than 1%. This result is important, because it shows the appropriateness of the dissipation-fluctuation mechanism we introduced in eq. (18).

| *General mprDPD-parameters in reduced units* | Value |
|---|---|
| Particle density water ρ (determining pressure) | 3 |
| Cut-off radius $r_C$ | 1 |
| $k_B T$ | 1 |
| time step $\Delta t$ | 0.003 |
| mass and moment of inertia of water/membrane part. | 1 |

| *Interaction param. cf. eqs* (9) **and** (11) | *water water* | *water-lipid* | *lipid-lipid* |
|---|---|---|---|
| $\gamma_{ww}, \gamma_{wb}, \gamma_{ll}$ | 4.5 | 4.5 | 4.5 |
| $F_{max}$ | 25 | 25 | 100 |
| $F_{min}$ | 0 | 0 | -20 |
| $r_{rf}$ | 1 | 1 | 0.7 |
| α | | | 100 |
| β | | | 20 |
| *b* (density dept. interact.) | | | 0.7 |

**Table 1**: Parameters used in the simulations. Values are given in DPD-units determined by choosing the length of the cut-off radius, fixing $k_B T$ and then equalizing particle and mass density.

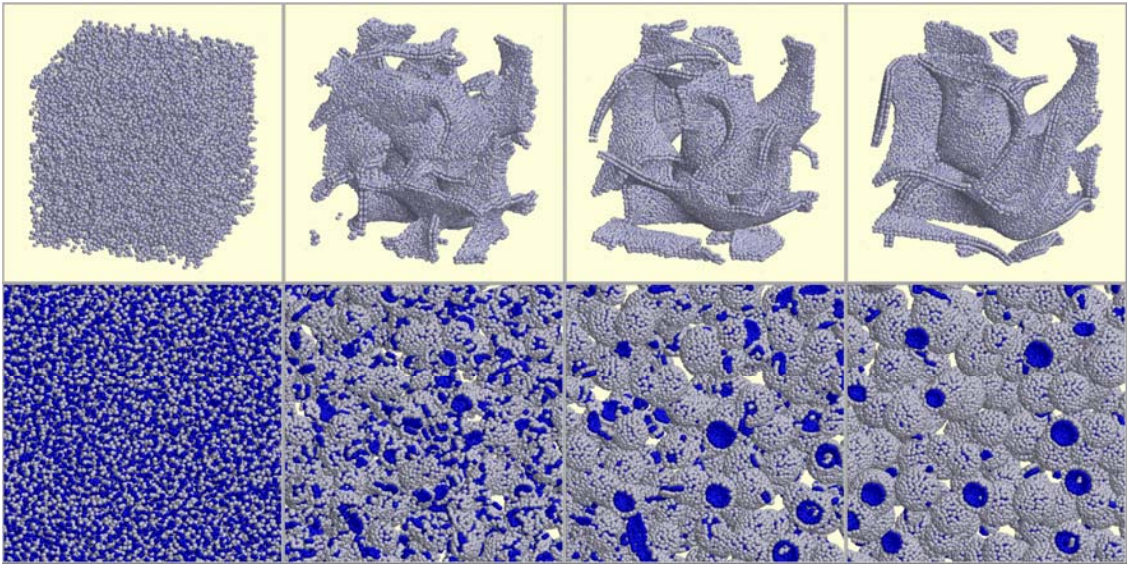

Fig. 3: **Self-assembly of different phases.** Top: self-assembled laminar sheets at t= 0, 0.6, 1.2, and 6 $\tau$ . Calculations were performed in a cube of side length $16 r_C$ with 10240 membrane particles and periodic boundaries. Bottom: self-assembled vesicles, partial view of a cube of side length $30 r_C$ with 67500 membrane particles. The number of water particles was dynamically adjusted in the transient phase in order to get a pressure equivalent to that in system containing only water and with a particle density of $\rho = 3$ . In case of the sheets, we set $\beta = \kappa = 0$ and for vesicles $\beta = 20, \kappa = 0.6$ . The simulations required approximately 12h on a conventional CPU.

**Statistical mechanics of extended objects**

A vesicle consisting of $n_{ves} = 220$ membrane particles is embedded into and filled with a solvent and trapped in a spherical harmonic potential of the form

$$U(r) = ar^2 .$$  (20)

The potential interacts with the membrane particles, and the parameter of the potential is set to $a = 0.01$ . This implies that, depending on the distance from the center of the potential, there is a total external force on the vesicle

$$\vec{F}_{ves} = -n_{ves} \nabla U = -a n_{ves} r .$$  (21)

The random forces lead to a fluctuation of the center of mass of the whole vesicle and an analysis of the trajectory of the vesicle is expected to yield a distribution for the distance $r$ from the center of the potential of the form:

$$\rho(r) = \frac{4r^2}{\pi} \sqrt{\frac{a^3}{k_B^3 T^3}} \exp(-\frac{ar^2}{k_B T}) \,. \tag{22}$$

Fig. 4 shows the comparison of the measured distribution in a simulation and compares it with the expected one. It turns out that the measured data would be optimally fitted by a distribution function of the above form with $T = 0.94$ (in units of the DPD simulation, for their interpretation in terms of conventional units see below) whereas the effective temperature in the simulation was set to $T = 1.0$. We conclude that, at least in the situation we investigated, the behavior of extended objects is well reproduced with respect to statistical mechanics. A better match of the temperature can be obtained by longer sampling.

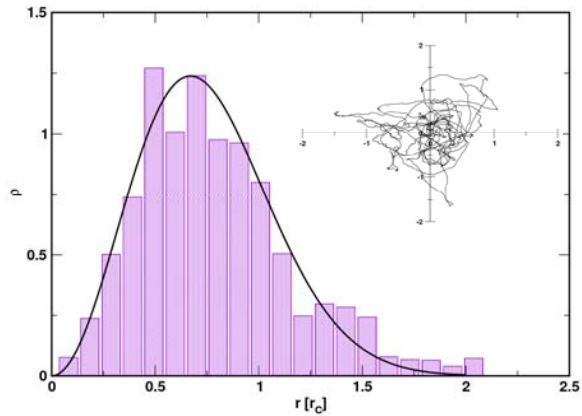

Fig 4: **A vesicle in a spherically symmetric harmonic potential.** The vesicle is composed of 220 membrane particles. In the inset, the trajectory is shown as a two-dimensional projection and in the main figure the distribution with respect to the distance to the center of the potential is plotted. The green curve gives the best-fit distribution, as expected from equilibrium statistical mechanics. This best fit corresponds to a distribution at T = 0.98 whereas the actual simulation was performed with a system temperature of T = 1.04, measured from particle velocities (Temperatures in units of the DPD-simulation, for the transformation into standard physical units see text.) This result provides an example of the correctly scaling transfer of statistical mechanical properties from locally interacting particles to extended objects.

## Emergent dynamics- Stokes' law

A more revealing test of the dynamics is provided by checking the validity of Stokes' law. The hydrodynamic behavior of extended objects in viscous fluids is a non-trivial collective problem. Pulling a spherical object in a viscous fluids has effects over a length scale of many diameters of the pulled object and the proper form of Stokes' law only emerges if this dynamics, which is far above the length scale of particle-particle interactions, is correctly reflected as an emergent phenomena on the mesoscale. Simulating the hydrodynamic behavior of vesicles is not only a benchmark from a theoretical point of view, but also an issue for the interpretation of present experimental investigations in cellular biology[26]. Stokes' law predicts that if a spherical object is pulled with a constant force of size $F$, this object will eventually attain a stationary velocity $v_{stat}$ which is given by

$$v_{stat} = \frac{F}{6\pi\eta r} \ .$$

(23)

Here, $\eta$ stands for the viscosity of the solvent. Fig. 5 shows that in fact, the stationary velocity of vesicles in the simulation is proportional to the inverse of the vesicle radius over a large range of different radii.

This result is not only relevant for the proof that mesoscale emergent dynamics is reproduced correctly by the simulation but also for another reason which has to do with the way the force was applied to the vesicle. Only a single membrane particle was subject to a pulling force. The force was properly distributed over the whole, much larger vesicle and of course also on to the aqueous contents of the vesicle. It is remarkable that although the vesicles are rather soft entities, the membranes themselves are fluid, the integrity of the whole vesicle was maintained and the particle being pulled was not torn out from the vesicle surface. This not only shows the advantages of a momentum conserving simulation method (which generically supports proper force transfer) but is also of relevance for biological simulations. For instance, moving a membrane bound structure along the cytoskeleton just by connecting an anchor with a motor protein is a basic task if one aims to understand dynamics and control mechanisms on the length scale of the whole cell.

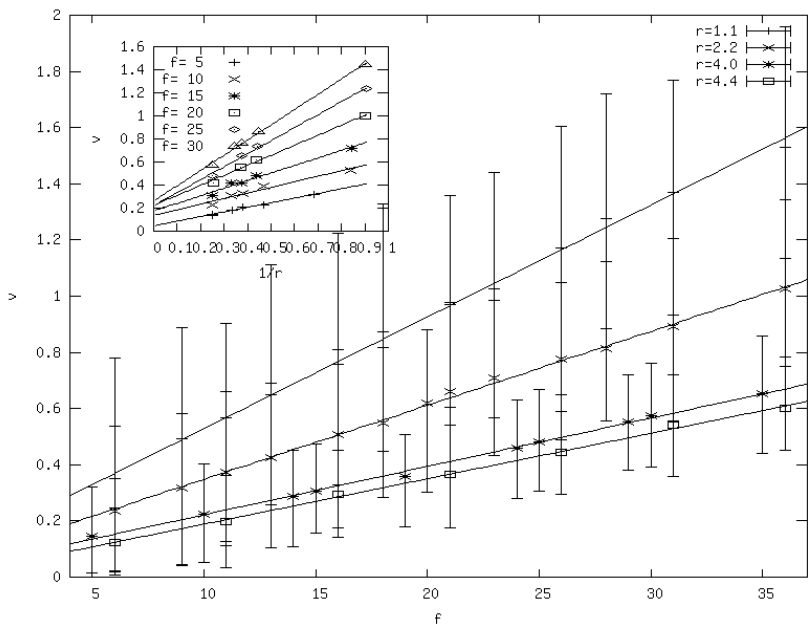

Fig 5: **Stokes' law:** the drift velocity of vesicles of varying radius $r$ is measured when subject to a constant force $f$ (shown for different forces). According to the laws of hydrodynamics, the vesicles experience a drag force that is proportional to their radius and consequently the vesicles will eventually attain a stationary velocity that is proportional to the inverse of the vesicle radius. That the stationary velocity follows a simple scaling law is due to an emergent dynamics on a length scale much greater than the interaction of individual particles in the simulation. The inset shows the measured velocity $v$ versus the inverse of the radius and shows that the mesoscale dynamics resulting from the mprDPD calculation does exhibit the correct scaling. Calculations were performed in a cube of side length 50 $r_c$ with 375000 particles. In the simulation, a vesicle is pulled by only exerting a force on one DPD particle, from among the several hundreds of them forming the complete vesicle. Taking into account that the membranes themselves are fluid and are held together only by local interactions of limited range (no springs or other stabilizing means), this shows that mprDPD is potentially capable of simulating transport processes of extended membrane bound compartments along the cytoskeleton.

## Discussion

The choice of a particle-based, coarse-grained approach, such as mprDPD has to be seen in the light of alternative methods. The combinatorially large number of possible molecular constituents makes it inappropriate to tackle the simulation of the dynamical chemical machinery of the cell using an approach based on partial differential equations in general. It is emphasized, however, that for a variety of specific questions, such as $Ca^{2+}$-waves [27] or even for molecular aggregation processes in defined geometries[28] for example, the PDE-approach can be both numerically efficient and provide analytically insight. By contrast, what we desire in this work is a general tool for mesoscale reactions and dynamics with the kind of generality provided by molecular dynamics, but for the timescales of cellular processes and dealing with chemical reactions.

Molecular dynamics[29] (MD) delivers the most detailed results, but presently on nanosecond to microsecond timescales, far below the requirements for simulating extended structures on timescales of seconds. Today, only patches of membrane can be simulated, but then also delivering information down to molecular detail[30]. Although in principle derivable from atomic details ab initio, good potentials for MD are derived or corrected semi-empirically, in order to match results to particular systems. In systems biology, interaction networks (genomic, proteomic, etc.) are also related to experimental (high throughput) data, even if it is clear that those networks are incomplete and some of the interactions are only quantified phenomenologically. Our mprDPD approach shares a phenomenological derivation of local interactions, but is aimed at being consistent for mesoscale modeling of cellular processes.

Macroscopic compartmentalized reaction kinetics is a widely used approach to model cellular systems, see for example the "Virtual Cell"[31]. Based on communicating, hierarchically organized compartments, these models represent the topology (and in some cases, not yet supported by standardized interfaces, the static geometry), but no other aspects of spatial organization of the cell. Models of this type have been standardized and implemented as versatile platforms that can read user-models in a communicable form: the Systems Biology Markup Language (SBML) is a widely accepted format for exchange and representation of specific applications[32]. However, compartment models cannot account generically for emergent phenomena of dynamic compartment formation. Besides technical difficulties in specifying dynamical compartments, SBML suffers from not treating the interplay of local interactions and global structural dynamics i.e. the physics of the cell. Membrane morphology for example is a genuine consequence of collective self-assembly processes and protein modulation in three dimensions and cannot be reduced to pure topological operations. In addition to these shortcomings, whether a compartmentalized kinetic model is formulated using sets of ordinary differential equations (ODE), or stochastic kinetics via master equations, it is difficult to handle combinatorially diverse families of molecules.

A further possibility to model spatially resolved cellular processes employs pre-defined structures and studies their dynamics using known transformation processes[33]. This approach includes the effects of spatial organization, but processes such as the fusion of membrane compartments have to be "put in by hand", which although adequate for specific problems reduces the versatility of the simulation platform and complicates the linkage with typical high-throughput data. Note that ideally a systems biology simulation platform should work with parameters that can be measured in an automated manner. Phenomenological rates e.g. describing the rates of fission of compartments as a function of chemical composition would (besides other considerable experimental problems) require combinatorially complex match with kinetic geometric data. For a systems biology approach, such a fission event should be related to local molecular densities, or local correlations between such densities, which are ascertainable by fluorescence imaging or similar techniques.

DPD and mprDPD are chosen from other approaches for their computational efficiency. Lattice gas or lattice Boltzmann approaches methods are burdened by the fact that it is difficult to simulate extended objects (especially if they are allowed to rotate) and their intrinsic breaking of Galilean invariance, which means that extended objects with a non-vanishing velocity are hard to simulate.

In this work, we only considered dipole interactions between particles. This gave us the possibility to construct particles with directional properties, in some sense comparable to Shillock and Lipowski's[8] chain particles. Of course, the method can directly be extended to higher multipole moments, and we expect this to be of value for the study of supramolecular self-assembly processes. The comparison with Shillcock and Lipowski is illuminating insofar as the two different approaches for going beyond structureless point particles can be clearly worked out. One way is to couple such structureless particles artificially to larger compounds. This method keeps the basic interactions simple and allows a high resolution but fixes a length scale and may be numerically costly. The other way, and the one we pursued, is to introduce more complex interactions, whereby we have chosen the mathematical method of multipole expansion to achieve this goal.

The potentials we use for our dipole interaction may appear arbitrary and more complex ones can easily be imagined. Note, however, that the expressions we have chosen are just the first terms of any potential involving dipoles satisfying the necessary symmetry conditions. The vesicles we simulated were constructed from a single layer of membrane particles. Bilayered structures would, besides a finer resolution, require the introduction of hydrophobic interactions. We got for micellar systems satisfactory results with

$$U_{hph}(\vec{d}_i, \vec{r}_{ij}) = \chi \Omega(r_{ij})(\vec{d}_i \cdot \vec{e}_{ij}) \tag{24}$$

where it is assumed that particle $i$ is of dipolar type and $j$ represents water. If the roles of the particles are changed, eq. (24) gets a minus sign.

The mprDPD method presented has to be seen in relation to the Fluid Particle Method (FPM) pioneered by Espanol[34]. The exchange of angular momentum was recognized there as being relevant for establishing hydrodynamics in the general case and the FPM method was thoroughly investigated and justified by establishing contact with statistical mechanics. We emphasize that the mprDPD-method presented in this work handles angular momentum and torque in a, in some aspects, more general manner. Whereas the FPM-model can be understood as simulating the dynamics of extended spherical objects interacting also via non-central dissipative forces (thereby accounting properly for friction), mprDPD departs from the assumption of spherical particles. The direct way to implement this departure would be to assume a detailed geometry of the particles under consideration and implementing the proper mechanical interactions. Such a mechanical approach, however, is prohibitively complicated and computationally time-consuming. By employing the Lagrangian formalism, it is possible to circumvent these difficulties; the (geometrically non-trivial) particles are represented as multipole expansions (in this work truncated at the dipole term) and their interactions are governed by a Lagrangian that is chosen in such a way that it exhibits the same symmetries as the structures one wants to study. Non-central forces do appear in mprDPD, but cannot directly be understood as accounting for hydrodynamic interaction, as is done in FPM. Extending the mprDPD-formalism by the FPM-interactions is therefore desirable.

In summary, the problem that extended (quasi-)two-dimensional structures such as membranes cannot be simulated by isotropically interacting point particles can be overcome by several strategies. Firstly, conventional DPD-particles can be coupled with springs to larger entities. Secondly, the point particles can be replaced by extended entities as it is done in FPM or the Voronoi-based dissipative particle dynamics method by De Fabritiis, Coveney and Flekkoy[35]. The third approach and that which is adopted here is mprDPD, in which the usage of point-particles is retained, but the particles exhibit structure by equipping them with multipole moments.

One may raise the question whether a simulation based on individual particles can describe properly a continuous distribution of material in a cellular environment. Rather than repeating the arguments for a particle based approach which we raised, we refer to the most widely used particle based method, namely Smoothed Particle Dynamics (SPH). For a review, consult Monaghan[36]. The large experience collected with this method is also of relevance for mprDPD. One must be aware of the fact that every computation derives the values of probably continuous variables from finitely many points. This holds for all particle-based methods, but also for methods that use a fixed grid in space. From this point of view, moving particles can be understood as moving grid points; it then becomes clear that particle-based methods have several advantages, especially with respect to mixed fluids or fluids with (mesoscopically) non-vanishing flow.

Simulating complex biological systems requires the combination of processes at the molecular length scale with mesoscopic phenomena. The fact that there are already simulation methodologies[37] that integrate MD calculations with DPD and FPM models makes us confident that this can also be achieved when using mprDPD.

## Summary

Enhancing conventional DPD by reaction mechanisms and multipolar interactions between the particles, resulting form the generic coarse graining process of DPD, enables the study of complex mesoscale entities such as membrane bound compartments. In order to get self-organized membranes, the interactions between particles must show more structure than can result from simple isotropic forces, independently of how complex the distance dependence may be. We have shown that the implementation of multipolar interactions – for membranes dipoles are sufficient – is a valid and efficient alternative to dedicated particle-assemblies such as the chainlike structures, as employed in various DPD-based studies of membranes.

DPD is a generically well-suited tool for the study of extended objects. We have shown that not only stationary structures, as in equilibrium phase diagrams, but also mesoscale dynamical properties emerge properly from completely local interactions. Tests such as the correct reproduction of Stokes' law not only yielded the correct scaling behavior but also showed that mprDPD is well suited for the study of intracellular processes, such as the directed transport of membrane bound compartments through locally applied forces. This situation occurs for example in the transport of endosomes along the cytoskeleton. Note that the extended structure that was transported in this work was a vesicle filled with solvent and bounded not by a rigid structure but by a fluid membrane.

The fact that even complex mechanical and structural processes can be simulated with entities interacting only via local interactions is of considerable relevance for studies in systems biology. Experimentally accessible are usually only local concentrations of molecules (e.g. via fluorescence imaging) or other local quantities. This means that an mprDPD-based platform can be directly linked up with high throughput imaging data and therefore can serve as tool that allows an automated connection of experiments with theoretical predictions, thereby opening new possibilities for control and analysis in biological research.

## Acknowledgements

This work was supported by the BMBF project Systems Biology of the Liver Grant # 3P3137 and by the EU integrated project FP6-IST-FET PACE #02035.

### References

1 Rothmann JE, Wieland FT (1996) Protein Sorting by Transport Vesicles. Science 272: 227-234.

2 Scheckmann R, Orci L (1996) Coat Proteins and Vesicle Budding. Science 271: 1526-1533.

3 Barlowe C (2000) Traffic COPs and the early secretory pathway. Traffic 1: 1371-1377.

4 Hoogerbrugge PJ, Koelman, JMVA (1992) Simulating microscopic hydrodynamic phenomena with dissipative particle dynamics. Europhys Lett 19: 155.

5 Espanol P, Warren P (1995) Statistical mechanics of Dissipative Particle Dynamics. Europhys Lett 30: 191.

6 Groot, RD Warren PD (1997) Dissipative particle dynamics: Bridging the gap between atomistic and mesoscopic simulation. J Chem Phys 107: 4423.

7 Warren PB (1998) Dissipative particle dynamics. Curr Op Colloid science 3: 620.

8 Shillcock JC, Lipowsky R (2002) Equilibrium structure and lateral stress distribution of amphiphilic bilayers from dissipative particle dynamics simulations. J Chem Phys 117: 3048.

9   Ortiz V, Nielsen SO, Discher DE, Klein ML, Lipowsky R, Shillcock J (2005) Dissipative Particle Dynamics Simulations of Polymersomes J Phys Chem B 109: 17708-17714.

10  Li DW, Liu XY (2005) Examination of membrane fusion by dissipative particle dynamics simulation and comparison with continuum elastic models.  J Chem Phys 122, 174909.

11  Jakobsen AF, Mouritsen OG Weiss M (2005) Close-up view of the modifications of fluid membranes due to phospholipase A2. J Phys: Condens. Matter 17:  S4015–S4024

12  Jena BP, Cho SJ (2002) The atomic force microscope in the study of membrane fusion and exocytosis. Atomic Force Microscopy in Cell Biology Methods in Cell Biology 68: 33-55.

13  Murray JW, Wolkoff AW (2003) Roles of the cytoskeleton and motor proteins in endocytic sorting. Advanced Drug Delivery Reviews 55: 1385-1403.

14  Vattulainen I, Karttunen M, Besold G, Polson JM (2002) Integration schemes for dissipative particle dynamics simulations: From softly interacting systems towards hybrid models. J Chem Phys 116: 3967.

15  Noguchi H, Takasu M (2002) Structural dynamics of pulled vesicles: A Brownian dynamics simulation. Phys Rev E. 65: 051907-1-7.

16  Gibson JB, Chen K (1999) The equilibrium of a velocity-Verlet type algorithm for DPD with finite time steps. J Mod Phys C 10: 241.

17  Turelli M (1977) Random Environments and Stochastic Calculus. Theor Popul Biol 12: 140-178.

18  Kumar PBS, Gompper G, Lipowsky R (2001) Budding Dynamics of Multicomponent Membranes. Phys Rev Lett. 86: 3911-3914.

19  Dawson. KA and Kurtovic, Z, (1990) Lattice model of self-assemlby. J. Chem. Phys. 92: 5473-5485.

20  Pagonabarraga I and Frenkel D (2001) Dissipative Particle Dynamics for Ineracting Systems. J. Chem. Phys. 115: 5015-5026.

21  Warren PB (2003). Vapor-liquid coexistence in many-body disspipative particle dynamics. Phys. Rev. E 68:066702.

22  Trofimov, SY, Nies ELF and Michels, MAJ (2002) Thermodynamic consistency in dissipative particle dynamics simulations of strongly non-ideal liquids and liquid mixtures. J. Chem. Phys. 117:9383-9394.

23  Groot RD Novel Methods in Soft Matter Simulations, volume 640 of Lecture Notes in Physics, chapter Applications of Dissipative Particle Dynamics. Springer, 2004.

24  Füchslin RM,  Eriksson A, Fellermann H and  Ziock, HJ (2007). Coarse-Graining and Scaling in Dissipative Particle Dynamics, submitted.

25  Yamamoto S and Hyodo S (2003). Budding and fission dynamics of two-component vesicles. J Chem Phys 118(17):7937–7943.

26  Hill DB, Plaza MJ, Bonin K, Holzwarth G (2004) Fast vesicle transport in PC12 neurites: velocities and forces. Eur Biophys J 33: 623–632.

27  Kupfermann R, Mitra PP, Hohenberg PC, Wang SS (1997) Analytical Calculation of intracellular Calcium Wave Characteristics. Biophys J 72: 2430 – 2444.

28  Del Conte Zerial P, Brusch L, Deutsch A (2005) A spatio-temporal model of Rab5-domain formation, preprint.

29  Frenkel D, Smit B (2002) Understanding Molecular Simulations: From Algorithms to Applications 2nd edition. Academic Press Inc. p. 664.

30  Goetz R, Lipowsky R (1998) Computer simulations of bilayer membranes: Self-assembly and interfacial tension. J Chem Phys 108: 7397.

31 Slepchenko BM, Schaff JC, Macara I, Loew, LM (2003) Quantitative Cell biology with the Virtual Cell. Trends Cell Biol 13: 570-576.

32 www.sbml.org

33 Gao J, Shi W, Freund LB (2005) Mechanics of receptor Mediated Endocytosis. Proc Natl Acad Sci U S A 102: 9469-9474.

34 Espanol P (1998) Fluid particle model. Phys Rev E 57, 2930-2948.

35 De Fabritiis G, Coveney PV, Flekkoy EG (2002) Multiscale dissipative particle dynamics. Phil Trans R Soc Lond A 360: 317–331.

36 Monaghan JJ (2005) Smoothed particle hydrodynamics. Rep Prog Phys 68: 1703-1759.

37 Dzwinel W, Yuen D, Boryczko K (2002) Mesoscoopic dynamics of colloids simulated with dissipative particle dynamics and fluid particle model. J Mol Model 8: 33-43.